\documentclass[article,11pt,showpacs,amsmath,amssymb]{revtex4-1}
\usepackage{graphicx}
\pdfoutput=1
\usepackage{amsmath}
\usepackage{color}
\usepackage{float}
\setcitestyle{super}
\usepackage[english]{babel}
\usepackage{color,soul}
\usepackage{color}
\usepackage[font=small,labelfont=bf, figurename=Fig.]{caption} 

\usepackage[pdftex]{hyperref}

\begin{document}
\title{Enhanced zero-bias conductance peak and splitting at mesoscopic interfaces between an $s$-wave superconductor and a 3D Dirac semimetal}

\author{Leena Aggarwal$^1$, Sirshendu Gayen$^1$, Shekhar Das$^1$, Gohil S. Thakur$^2$, Ashok K. Ganguli$^{2,3}$ \& Goutam Sheet$^1$}

\affiliation{$^{1}$Department of Physical Sciences, Indian Institute of Science Education and Research(IISER), Mohali, Sector 81, S. A. S. Nagar, Manauli, PO: 140306, India. \\$^{2}$Department of Chemistry, Indian Institute of Technology, New Delhi 110016, India. \\$^{3}$Institute of Nano Science $\&$ Technology, Mohali 160064, India.}

\begin{abstract}

\textbf{Mesoscopic point contacts between elemental metals and the topological 3D Dirac semimetal Cd$_3$As$_2$ have been recently shown to be superconducting with unconventional pairing while Cd$_3$As$_2$ itself does not superconduct. Here we show that the same superconducting phase at mesoscopic interfaces on Cd$_3$As$_2$ can be induced with a known conventional superconductor Nb where a pronounced zero-bias conductance peak is observed which undergoes splitting in energy under certain conditions. The observations are consistent with the theory of the emergence of Andreev bound states (ABS) due to the presence of a pair potential with broken time reversal symmetry. The data also indicate the possibility of Majorana bound states as expected at the interfaces between $s$-wave superconductors and topologically non-trivial materials with high degree of spin-orbit coupling.} 
\end{abstract}

\maketitle
\makeatletter
\renewcommand\@biblabel[1]{#1}
With the discovery of every complex superconducting phase, particularly in the newly discovered topologically nontrivial systems, certain most important fundamental questions related to the nature of superconductivity must be addressed before the mechanism of unexpected superconductivity in such nontrivial systems can be understood. On the other hand, understanding of induced superconductivity in the topological systems also hold the promise of ultimately realizing the topological superconductor that are expected to host some of the most elusive particles predicted in quantum field theory, the Majorana fermions\cite{Frank1, Frank2} which, due to their non-Abelian statistical properties are potentially important for fault-tolerant quantum computing\cite{Cheng,Clarke}. Recently, the 3D Dirac semimetal Cd$_3$As$_2$ \cite{Liu1, Liu2, Young1} emerged as such a material hosting Dirac fermions in the bulk where the material itself does not show superconductivity under normal conditions but nano-meter scale interfaces between normal metals and Cd$_3$As$_2$ become superconducting with unconventional pairing interactions\cite{Goutam}. This phenomenon has been categorized as tip induced superconductivity (TISC). A TISC phase on Cd$_3$As$_2$ was also confirmed by other groups\cite{Jian}. In this paper we report transport spectroscopic measurements on mesoscopic junctions between a well known conventional $s$-wave superconductor Nb and the 3D Dirac semimetal Cd$_3$As$_2$. The key observations include a pronounced zero-bias conductance peak which undergoes splitting as the effective barrier potential at the interface is physically tuned. The observations are consistent with the theory of the emergence of Andreev bound states (ABS) due to the presence of a pair potential with broken time reversal symmetry \cite{Tanaka1,Golu,Cov}. The data also indicate the possibility of Majorana bound states as expected at the interfaces between $s$-wave superconductors and topologically non-trivial materials with high degree of spin-orbit coupling\cite{Ando,Hui2,Cole2,Satiawan,Cole1,Takei,Liu,Stan1,Lut,Kim,Hui1,Ganeshan,Stan2}.

\begin{figure*}[h]

	\includegraphics[scale=0.65]{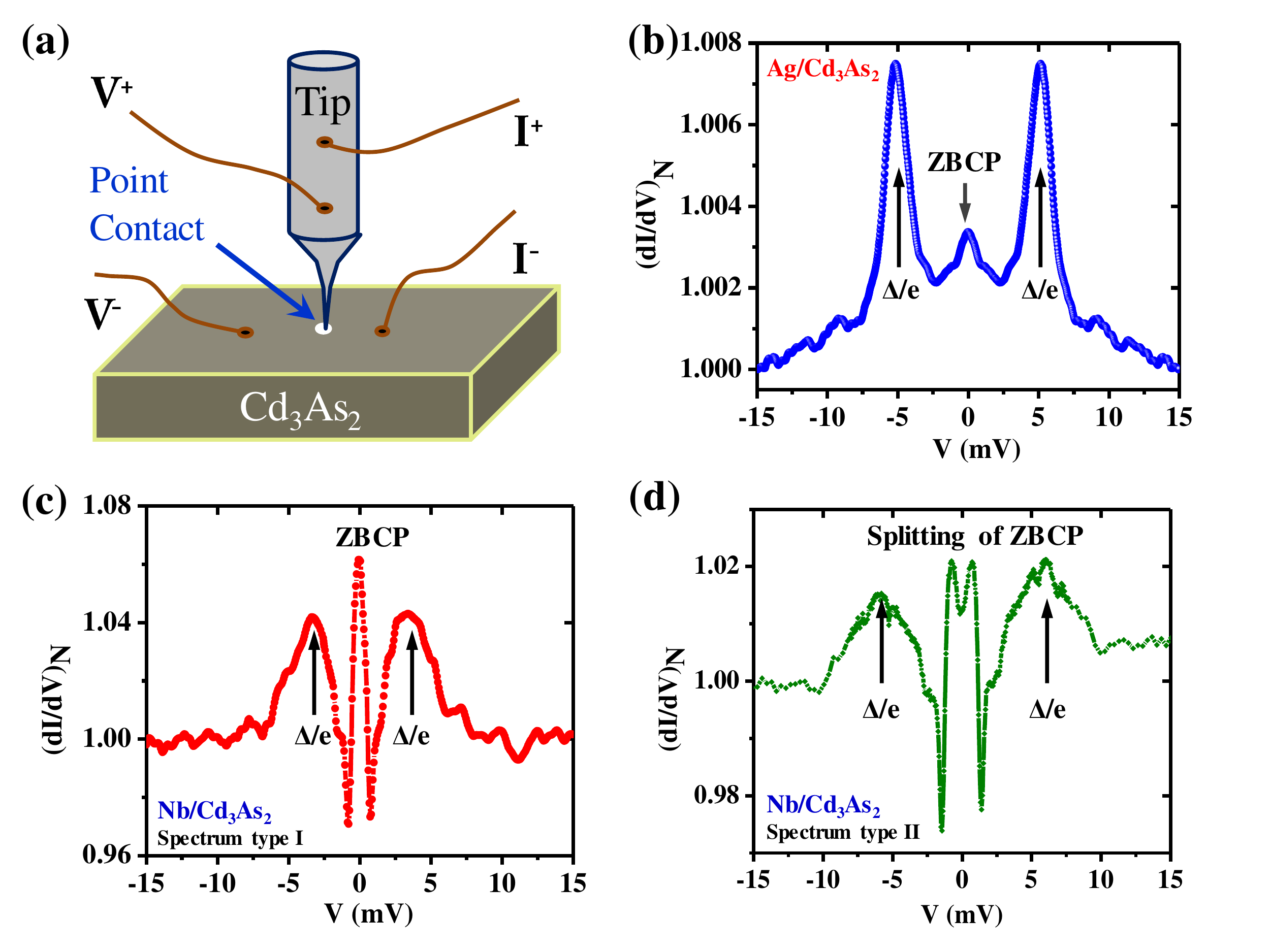}
	\caption{(a) The schematic diagram showing the point-contact and the measurement electrodes. A normalized differential conductance ($(dI/dV)_N$) spectrum for (b) Ag/Cd$_3$As$_2$ point-contact showing a zero-bias conductance peak (ZBCP) along with two sharp peaks corresponding to the induced superconducting gap voltage. (c) Nb/Cd$_3$As$_2$ point-contact showing a very prominent ZBCP without splitting --- we call this type-I spectrum. (d) For a different barrier strength the ZBCP shows splitting --- we call this type-II spectrum.} 
\end{figure*}

In Figure 1(a) we show a schematic diagram of the experimental set up. A detailed description of the measurement technique is included in the supplementary materials. Figure 1(b) shows a typical point-contact spectrum obtained between a non-superconducting elemental metal like silver (Ag) and Cd$_3$As$_2$. All the spectral features seen here were discussed in detail in Ref. [1]. The two peaks symmetric about $V = 0$ in this spectrum appear due to Andreev reflection between the normal metal and a superconducting phase that is induced at the point of contact and the position of the peaks gives an approximate estimate of the superconducting energy gap which in this case was $\sim 5.5\,meV$. A tiny zero-bias conductance peak (ZBCP) is also observed in the spectrum. In Ref. [1] it was mentioned that this ZBCP could be a signature of a zero energy Majorana mode. However, since a ZBCP might originate in a point-contact spectrum from a large number of other physical phenomena that might take place at the points of contact, in Ref. [1] the observed ZBCP could not be claimed to be a conclusive signature of a Majorana mode.

Since from theoretical perspective significant progress has been made in understanding the possible ``smoking gun" signature of Majorana modes that might emerge at the interfaces between a known $s$-wave superconductor and a high spin-orbit coupling semiconductor\cite{Sarma}, we have performed point contact spectroscopy on Cd$_3$As$_2$ with sharp tips of Nb, a conventional $s$-wave superconductor and explored the evolution of the ZBCP as a function of temperature, magnetic field and the interfacial properties of the point contact. In Figure 1(c) we show a spectrum obtained from an Nb/Cd$_3$As$_2$ point contact with a normal state (high-bias) contact resistance of $2$ $\Omega$.The spectrum shows a superconducting gap structure at $\Delta =\pm 4.5\, meV$. This is slightly smaller than the gap ($\Delta = \pm 5.5\, meV$) observed for Ag/Cd$_3$As$_2$ point contacts as shown in Figure 1(b). This difference might be attributed to a non-trivial mixing of the $s$-wave gap of niobium with the unconventional gap induced on Cd$_3$As$_2$. Furthermore, in contrast to Figure 1(b), the spectrum presented in Figure 1(c) shows a remarkably sharper ZBCP. When the point contact is physically changed in order to achieve a different effective interfacial barrier ($Z$, as defined in the BTK theory),  the ZBCP spontaneously splits into two peaks symmetric about $V = 0$ (See Figure 1(d)). For all the experiments presented here, the Nb tips were made out of $0.25\, mm$ dia. wires of pure (99.999\%) Nb which shows a superconducting critical temperature of 9.1\,K and a gap of $1.2\, meV$. It should be noted that the spectra for Nb/Cd$_3$As$_2$ point contacts are broader than the spectra for Ag/Cd$_3$As$_2$ point contacts. This can be attributed to the fact that the 
$s$-wave gap of Nb mixes non-trivially with the unconventional gap of Cd$_3$As$_2$ and the Andreev reflection feature associated with Nb is hidden in the broadened spectra. 

\begin{figure*}[h]

	\includegraphics[scale=0.65]{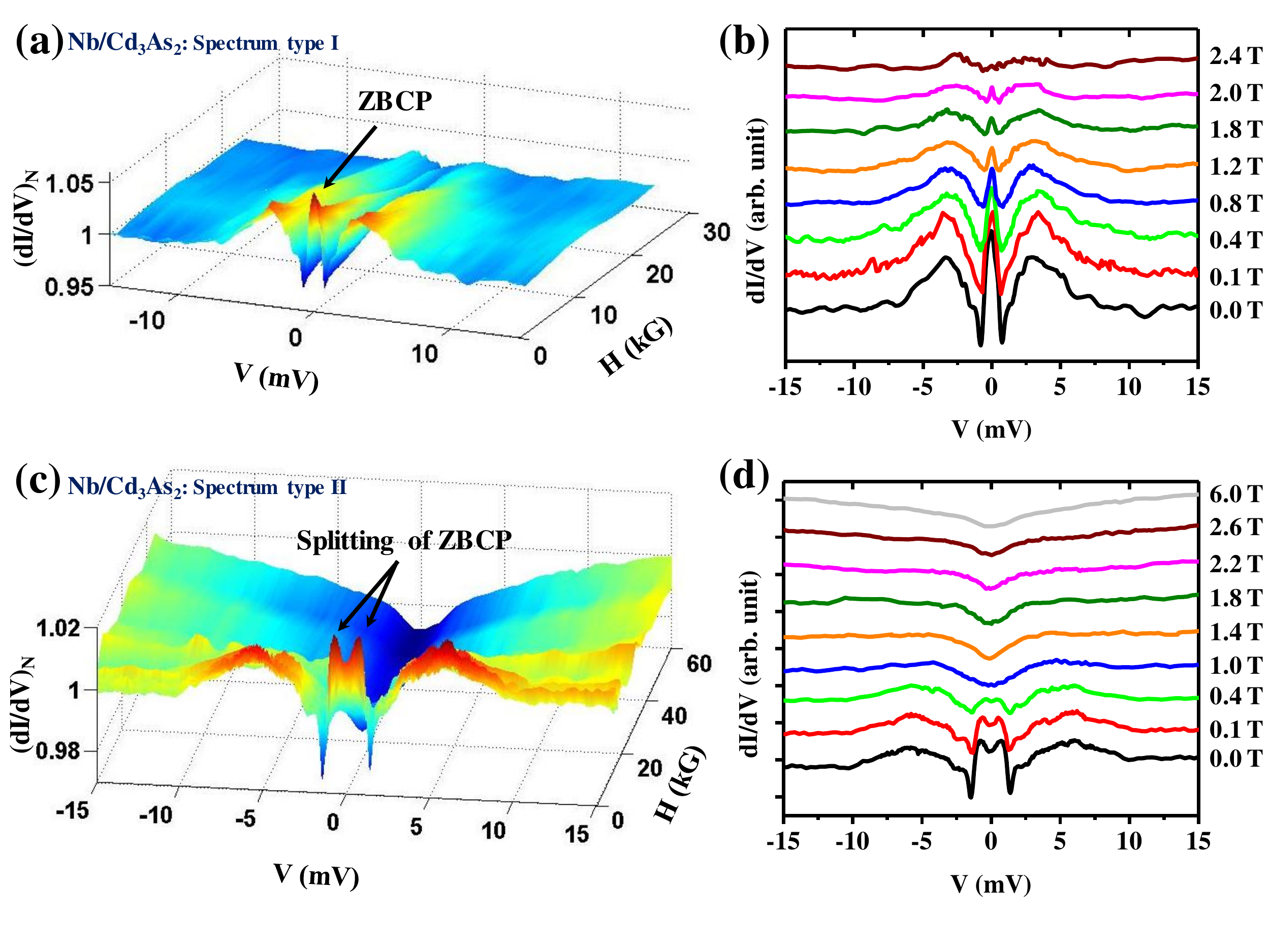}
	
	\caption{(a),(b) Magnetic field dependence of type-I spectra for Nb/Cd$_3$As$_2$ point-contact in 3-D and 2-D respectively. (c),(d) Magnetic field dependence of type-II spectra for Nb/Cd$_3$As$_2$ point-contact in 3-D and 2-D respectively. The point-contact spectra showing disappearance of a ZBCP and splitting of the ZBCP with stronger magnetic field.}
\end{figure*}

In order to find out whether the zero-bias conductance peak is associated with the superconductivity of the point contacts, we have performed magnetic field dependence of both types of spectra. The systematic magnetic field dependent data are presented in Figure 2. Both types of spectra, type I and type II evolve smoothly with magnetic field. The spectra of type I lose the prominent spectral features namely the induced gap structure at 5\,meV and the single ZBCP at a magnetic field of 2.4\,T (Figure 2a, Figure 2b). It should be noted that the ZBCP disappears at the same magnetic field where the induced superconducting gap structure disappears. Therefore, it can be concluded that the observed ZBCP is related to superconductivity. In fact, a similar magnetic field dependence of the ZBCP was seen with non-superconducting tips\cite{Goutam}. The spectra of type II that show the splitting of the ZBCP also show a systematic evolution with increasing magnetic field. The double peak structure gives rise to a single peak at 0.4 T which further evolves to become a single dip at 1.0\,T. The dip structure, along with the induced gap structure at higher bias, almost disappears at a high magnetic field of 6 T. The difference in the magnitude of fields at which the spectral features disappear for the spectra of type I and of type II further confirm that the nature of the point contacts in the two cases are significantly different. The observed difference in magnetic field dependence for the two different point contacts is consistent with the expected variation in superconducting point contacts.

The relation between the ZBCP and it's splitting can also be confirmed from the temperature dependence of the Andreev reflection spectra of type I and type II (Figure 3). All the spectral features evolve systematically with increasing temperature and the ZBCP disappears at the same temperature where the superconducting gap structure also disappears. It may be noted that the critical temperature of the point contacts vary slightly between the spectra of type I and of type II possibly due to the difference in the point contact geometry and the transparency of the point contacts.

\begin{figure*}[h]
	
	\includegraphics[scale=0.65]{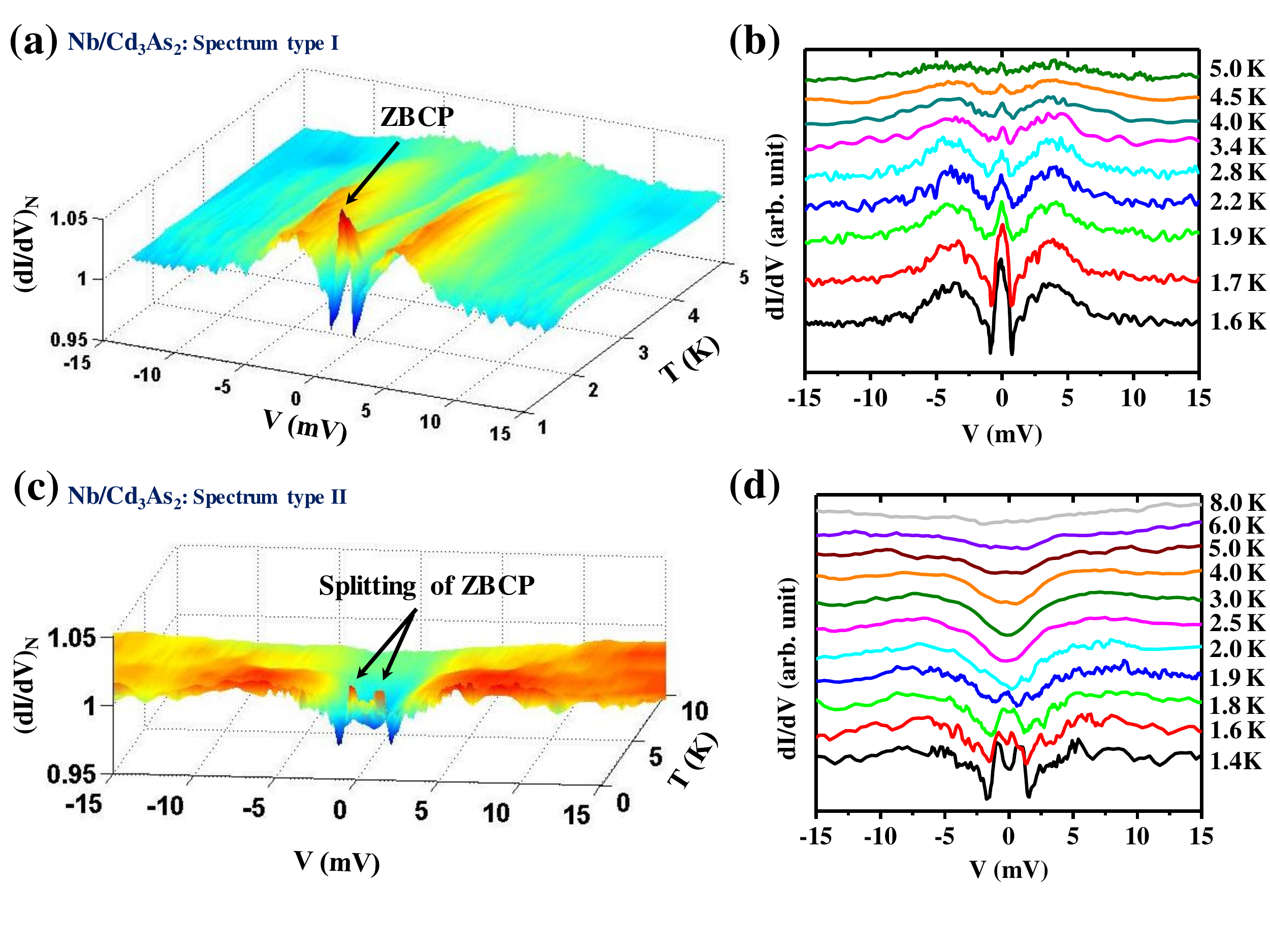}
	
	\caption{(a),(b) Temperature dependence of type-I spectra for Nb/Cd$_3$As$_2$ point-contact in 3-D and 2-D respectively. (c),(d) Temperature dependence of type-II spectra for Nb/Cd$_3$As$_2$ point-contact in 3-D and 2-D respectively. The point-contact spectra showing disappearance of a ZBCP and splitting of the ZBCP with higher temperature.}
\end{figure*}

\newpage
Having confirmed that the observed ZBCP is due to the superconducting nature of the point contacts, we concentrate on the significance of the ZBCP and the variation of the same with temperature and magnetic field respectively. Traditionally it has been believed that in point contact spectroscopy the ZBCP might arise when an opposite phase is encountered by the incident electrons and the Andreev reflected holes due to the presence of an unconventional pair potential in the superconductor and consequent emergence of the Andreev bound states (ABS) \cite{Tanaka1,Golu,Cov}. This is possible in Nb/Cd$_3$As$_2$ point contacts as well since it is known that the induced superconducting phase in the Cd$_3$As$_2$ point contacts is unconventional in nature. In such a scenario, when the time reversal symmetry is broken, for example by the application of an external magnetic field, the ZBCP undergoes splitting. However, as it is seen in our case, for the point contacts of type II, the ZBCP undergoes splitting spontaneously, even in zero magnetic field. This means, there is a internal mechanism present which facilitates the breaking of time reversal symmetry within the system even in absence of an externally applied magnetic field. 

In order to identify the possible mechanism of time reversal symmetry breaking in the Cd$_3$As$_2$ point contacts, we first discuss certain other systems where similar observations were made. In the context of the cuprate superconductors it was shown that for low level of doping the $dI/dV$ spectra exhibits a single ZBCP. The ZBCP undergoes splitting for samples with higher doping\cite{Dagan}. In this case the spontaneous splitting of the ZBCP was attributed to the presence of a complex $id_{xy}$-like component in the order parameter. In case of normal metal/unconventional superconductor junctions it was theoretically shown that the ZBCP can undergo splitting in zero magnetic field due to scattering from the impurities at the interfaces\cite{Asano}. Within this model the role of the interfacial impurity is simply to suppress the conductance at zero-bias thereby leading to the observed splitting. However, in both the cases discussed above, there is a unique dependence of the splitting predicted for the ZBCP. The splitting width is expected to increase with increasing magnetic field which is opposite to what we observe here (Figure 2(c) and Figure 2(d)). Therefore, these possibilities can be ruled out for the observed ZBCP splitting in Nb/Cd$_3$As$_2$ point contacts. It was also shown that a ZBCP and the spontaneous splitting of the same can be observed in tunneling data on a superconductor where the superconducting condensate is formed in multiple bands crossing the Fermi surface and the splitting of the ZBCP can be induced by tuning the interband coupling \cite{Pasanai}. However, at this point, due to the absence of theoretical understanding of the bands from which superconductivity might emerge in Cd$_3$As$_2$ point contacts, this picture cannot be tested.

The most consistent mechanism of the appearance of a ZBCP and the splitting of the same seems to be the presence of a spin-triplet component in the order parameter symmetry where the time-reversal symmetry is intrinsically broken in the system \cite{Tanaka}. The possibility of the existence of a spin-triplet component was clearly shown in the Andreev reflection data with a spin polarized cobalt tip where Andreev reflection was seen to be more pronounced than in the point contacts with non-ferromagnetic tips \cite{Goutam}. Furthermore, the unusual robustness of the ZBCP with increasing magnetic field up to a large field strength of 2.4\,T also hints to the possibility of time-reversal invariant Majorana edge modes in the Cd$_3$As$_2$ point contacts. Strikingly similar observation was made for point contacts on the topological superconductor Cu$_x$Bi$_2$Se$_3$ where the observation was attributed to the existence of zero-energy Majorana modes\cite{Sasaki}.  Within this picture the understanding of the spontaneous splitting of the ZBCP also emerges naturally. Tanaka $et al.$ had earlier shown that for ballistic junctions on superconductors with a spin triplet component in the superconducting order parameter, depending on the transparency of the junction either a single ZBCP (low $Z$) or a spontaneously split (high $Z$) ZBCP might emerge \cite{Yama}. Our observation is consistent with this description as for different point contacts giving rise to the two types of spectra shown in Figure 1(c) and Figure 1(d), the corresponding values of point contact $Z$ are different. 

In conclusion, we have reported a robust ZBCP and the spontaneous splitting of the same when the transparency of the point contacts is tuned in the differential conductance spectra obtained on the point contact induced superconductivity in the 3D Dirac semimetal Cd$_3$As$_2$ using a tip made of Nb, a conventional BCS superconductor. The results are consistent with the emergence of zero-energy Andreev bound states due to the presence of a time reversal symmetry broken pairing potential. The results also indicate the possibility of time reversal symmetry breaking Majorana edge modes in the point contact induced superconducting phase.  

\newpage
 
\textbf{Supplementary Material:}

The information related to synthesis, characterization and  additional spectroscopic data have been included in supplementary materials.

\textbf{\textbf{\textbf{Acknowledgement:}}}\\\
GS would like to acknowledge partial financial support from the research grant of Ramanujan fellowship awarded by the department of science and Technology (DST) govt. of India under the grant number SR/S2/RJN-99/2011 and the research grant from DST-Nanomission under the grant number SR/NM/NS-1249/2013.

%

	

\end{document}